\documentstyle[12pt]{article}
\date {}
\begin{document}

\centerline{ \bf CHIRAL SYMMETRY AND THE CONSTITUENT QUARK MODEL}

\vspace{1cm}
\centerline{L. Ya. GLOZMAN }

\vspace{0.2cm}
\centerline{\it Institute for Theoretical Physics,
University of Graz, 8010 Graz, Austria}

\centerline{\it Alma-Ata Power Engineering Institute,
480013 Alma-Ata, Kazakhstan}

\begin{abstract}
 New results on baryon structure
and spectrum developed in collaboration with Dan Riska
 \cite{GLO1,GLO2,GLO3,GLO4} are reported.
The main idea is that beyond the chiral
symmetry spontaneous breaking scale light and strange baryons
should be considered as  systems
of three constituent quarks with an effective confining interaction and a
chiral interaction that is mediated by the octet of Goldstone bosons
(pseudoscalar mesons) between the constituent quarks.
\end{abstract}

\vspace{0.5cm}

{\bf 1. Why does the Gluon Exchange Bear no Relation
to Baryon Spectrum}

\vspace{0.5cm}
It was accepted by many people (but not by all) that the fine splittings
in the baryon spectrum are due to the gluon-exchange interaction
between the constituent quarks \cite {DERU,IGK1}. Now I shall only
address to formal consideration of why the one gluon exchange interaction
cannot be relevant to the baryon spectrum.

\medskip
The most important component of the one gluon exchange interaction
\cite {DERU} is so called color-magnetic interaction

$$H_{cm}\sim-\alpha_s\sum_{i<j} \frac {\pi}{6m_im_j} \vec \lambda_i^C\cdot
\vec \lambda_j^C \vec
\sigma_i\cdot \vec \sigma_j\delta(\vec r_{ij}),\eqno(1.1)$$

\noindent
where the $\{\vec \lambda_i^C\}$:s are color $SU(3)$ matrices.
It is the permutational color-spin symmetry of the 3q state
which is mostly responsible for the contribution of the interaction
(1.1). The corresponding two-body matrix element is

$$<[f_{ij}]_C\times [f_{ij}]_S:[f_{ij}]_{CS}|\vec \lambda^C_i \cdot \vec
\lambda_j^C
\vec\sigma_i \cdot \vec \sigma_j|[f_{ij}]_C \times [f_{ij}]_S:[f_{ij}]_{CS}>
=\left\{\begin{array}{rr}  8 &
[11]_C,[11]_S:[2]_{CS} \\  -{8\over
3} & [11]_C,[2]_S:[11]_{CS}\end{array}\right..\eqno(1.2)$$

\noindent
Thus the symmetrical color-spin pairs (i.e. with the $[2]_{CS}$
Young pattern) experience an attractive contribution while the
antisymmetrical ones ($[11]_{CS}$) experience a repulsive contribution.
Hence the color-magnetic contribution to the $\Delta$ state ($[111]_{CS}$)
is  more repulsive than to the nucleon ($[21]_{CS}$) and
the $\Delta$ becomes heavier than the nucleon. The prise is that $\alpha_s$
should be larger than unity, which is  bad.

\medskip
 This interaction does not
practically contribute to the $N(1535)-N$ splitting as both these states have
identical mixed color-spin symmetry. Hence this splitting within this
model should be due to the spin-independent confining forces which means
that $\hbar \omega \simeq 500 - 600$ MeV. This large value of the harmonic
oscillator parameter implies a very small value for the nucleon radius,
$\sqrt{<r^2>}=\sqrt{\hbar/m\omega}\simeq 0.5$ fm,
if the light quark constituent mass is taken to be 330-340 MeV,
as suggested by the magnetic moments of
the nucleon.

\medskip
The crucial point is that the interaction (1.1) cannot explain
different ordering of the positive and negative parity states
in the spectra of the nucleon and the $\Delta$ resonance on the one hand,
and the $\Lambda$ - hyperon on the other. Indeed, the positive
parity state $N(1440)$ and the negative parity one $N(1535)$ have
the same mixed ($[21]_{CS}$) color-spin symmetry thus the color-magnetic
contribution to these states cannot be very different. But the $N(1440)$
state belongs to the $N=2$ shell while the $N(1535)$ resonance is
a member of the N=1 band which means that the $N(1440)$ should lie
approximately $\hbar \omega$ above the $N(1535)$. In the $\Delta$ spectrum
the situation is even more dramatic. The $\Delta(1600)$ positive parity
state has completely antisymmetrical CS-Young pattern ($[111]_{CS}$),
while the negative parity state $\Delta(1700)$ has the mixed one. Thus
the color-magnetic contribution to the $\Delta(1600)$ is much more
repulsive than to the $\Delta(1700)$. In addition the $\Delta(1600)$ is the N=2
state while the $\Delta(1700)$ belongs to the N=1 band.
As a consequence the $\Delta(1600)$ must lie much higher
than the $\Delta(1700)$.
In the spectrum of the $\Lambda$-hyperon
on the other hand it is the negative parity states $\Lambda(1405)-
\Lambda(1520)$ that remain the lowest lying resonances.

\medskip
Finally, there is no empirical indications in the spectrum
for the large spin-orbit
component of the gluon-exchange interaction \cite {DERU}.\\

{\bf 2. Chiral Symmetry and the Quark Model}
\vspace{0.5cm}

It is well known that at low temperature and density the approximate
chiral symmetry of QCD is realized in the hidden Nambu-Goldstone mode.
The hidden mode of chiral symmetry
is revealed by the existence of the octet of  pseudoscalar
mesons of low mass, which represent the associated approximate
Goldstone bosons. The $\eta'$ (the $SU(3)$-singlet) decouples from the original
nonet because of the $U(1)$ anomaly \cite{WE,THO}.
Another consequence of the spontaneous breaking of the approximate
chiral symmetry of QCD is that the valence quarks acquire
their dynamical or constituent mass
\cite{WEIN,MAG,SHU,DIP}
through their interactions with the collective excitations of
the QCD vacuum-
the quark-antiquark excitations and the instantons.

\medskip
We have recently suggested \cite{GLO1,GLO2}
that beyond the chiral symmetry
spontaneous breaking scale a baryon should be considered
as a system of three constituent quarks
with an effective quark-quark interaction that is formed
of a central confining part, assumed for simplicity to be harmonic,
and a chiral
interaction that is mediated by the octet of pseudoscalar mesons
between the constituent quarks.
	The simplest representation of the most important
component of the interaction of the constituent quarks that is mediated
by the octet of pseudoscalar bosons in the $SU(3)_F$
invariant limit is

$$H_\chi\sim -\sum_{i<j}V(\vec r_{ij})
\vec \lambda^F_i \cdot \vec \lambda^F_j\,
\vec
\sigma_i \cdot \vec \sigma_j.\eqno(2.1)$$
Here the $\{\vec \lambda^F_i\}$:s are flavor $SU(3)$ Gell-Mann
matrices and the
$i,j$ sums run over the constituent quarks.

\medskip
 There is very good analogy to solid state physics.
As soon as we talk about dynamical objects - constituent
quarks - we should forget about original QCD degrees of freedom
(gluons and instantons) like
in the solid state  we describe the electric and thermo properties
of metals in terms of heavy dynamical electrons (cf. constituent quarks),
phonons, which are Goldstone excitations in the lattice (cf. pseudoscalar
mesons), and electron-phonon interaction (cf. constituent quark - pseudoscalar
meson vertex). The one gluon exchange or instanton-induced interactions
between the constituent quarks
would correspond to the Coulomb interaction between the heavy
electrons in the lattice which is known to be totally unessential.

\medskip
The importance of the constraints posed by chiral symmetry
for the quark bag \cite{Chthorn} and bag-like \cite{Weise}
 models for the baryons
 was recognized early on .
 In the
bag or bag-like models
with restored chiral symmetry the
massless current quarks within the bag
were assumed to interact not only by
perturbative gluon exchange but also through chiral meson field exchange.
In these models
the chiral field  has the character of a compensating auxiliary field only
rather than a collective low frequency Goldstone quark-antiquark
excitation (the possibility of a nonzero quark condensate
was not addressed). A general limitation of all bag and bag-like models is
of course the lack of translational
invariance, which is important for a realistic description of the
excited states.

\medskip
Common to these models is that the breaking of chiral symmetry
arises from the confining interaction.
This point of view contrasts with that of Manohar and Georgi
\cite{MAG}, who pointed out that there should
be two different scales in QCD, with 3 flavors. At the first one
of these,
$\Lambda_{\chi SB} \simeq 4 \pi f_\pi \simeq$ 1 GeV, the
spontaneous breaking of the chiral symmetry occurs, and hence
at distances
beyond $ \frac {1}{\Lambda_{\chi SB}} \simeq$ 0.2 fm the valence
current quarks acquire their dynamical (constituent) mass
(called "chiral quarks" in \cite{MAG}) and the Goldstone bosons
(mesons) appear. The other scale, $\Lambda_{QCD}
\simeq 100-300$ MeV,
is that which characterizes confinement, and the inverse of this
scale roughly coincides with the linear size of
a baryon. Between these two scales then the effective Lagrangian
should be formed out of the gluon fields that provide
a confining mechanism as well as of the
constituent quark and pseudoscalar meson fields. Manohar and Georgi
did not, however, specify whether the baryons should be
desrcibed as bound qqq states or as chiral solitons.

\medskip
The chiral symmetry breaking scale above fits well with that which
appears in the instanton liquid picture of the QCD vacuum
\cite{SHU,DIP}. In this model the quark condensates (i.e. equilibrium
of virtual quark-antiquark pairs in the vacuum state)
as well as the gluon condensate
are supported by instanton fluctuations of a size $\sim 0.3$ fm.
Diakonov and Petrov \cite {DIP} suggested that at low momenta
(i.e. beyond the chiral symmetry breaking scale) QCD should
be approximated by an effective chiral Lagrangian of the sigma-model
type that contains valence quarks with dynamical (constituent)
masses and meson fields. They considered
a nucleon as three constituent quarks moving independently
of one another
in a self-consistent chiral field of the hedgehog
form \cite{DIAP}. In this picture the excited baryon states appear
as rotational
excitations and no explicit confining interaction is included.
A very similar description for the nucleon was suggested within so called
"chiral quark models" \cite{KAH,BIR}.\\

{\bf 3. The Chiral Boson Exchange Interaction}

\vspace{0.5 cm}

In an effective chiral symmetric description of baryon
structure based on the constituent quark model the
coupling of the quarks and the pseudoscalar Goldstone
bosons will (in the $SU(3)_F$ symmetric approximation) have
the form $ig\bar\psi\gamma_5\vec\lambda^F
\cdot \vec\phi\psi$, where $\psi$ is the fermion constituent quark
field operator and $\vec\phi$ the octet boson field
operator, and g is a coupling constant. A coupling of this
form in a nonrelativistic reduction for a constituent quark spinors
will -- to lowest order -- give rise to a Yukawa interaction
between the constituent quarks, the spin-spin component of which has
the form
$$V_Y (r_{ij})= \frac{g^2}{4\pi}\frac{1}{3}\frac{1}{4m_im_j}
\vec\sigma_i\cdot\vec\sigma_j\vec\lambda_i^F\cdot\vec\lambda_j^F
\{\mu^2\frac{e^{-\mu r_{ij}}}{ r_{ij}}-4\pi\delta (\vec r_{ij})\}
.\eqno(3.1)$$
Here $m_i$ and $m_j$ denote masses of the interacting quarks
and $\mu$ that of the meson. There will also be an associated
tensor component, which is discussed in ref. \cite{GLO3}.

\medskip
At short range the simple form (3.1) of the chiral boson exchange
interaction cannot be expected to be realistic, and should only
be taken to be suggestive.
Because of the finite spatial extent of both the constituent
quarks and the pseudoscalar mesons
that the delta function in (3.1) should be replaced by a finite
function, with a range of 0.6-0.7 fm as suggested
by the
spatial extent of the mesons.
In addition the radial behaviour of the Yukawa
potential (3.1) is valid only if the boson field
satisfies linear Klein-Gordon equation. The chiral symmetry
requirements for the effective chiral Lagrangian (which in fact is not known),
which contains
constituent quarks as well as boson fields
imply that these boson fields cannot be described by linear
equations near their source. Therefore it is only at large
distances where the amplitude of the boson fields is small that
the quark-quark interaction reduces to the simple Yukawa form.
{\it At this stage the proper procedure should be to avoid further specific
assumptions about the short range behavior of
$V(r)$ in
(2.1) and instead to extract the required matrix elements of it
from the baryon spectrum and to reconstruct by this an approximate
radial form of $V(r)$.}
The overall -- sign in the
effective chiral boson interaction in (2.1) corresponds to that of this
short range term in the Yukawa interaction.

\medskip
The flavor structure of the pseudoscalar octet exchange interaction
in (2.1) between two quarks i and j should be understood as
follows

$$V(r_{ij}) \vec {\lambda^F_i} \cdot \vec {\lambda^F_j}
\vec\sigma_i\cdot\vec\sigma_j =
\left(\sum_{a=1}^3 V_{\pi}(r_{ij}) \lambda_i^a \lambda_j^a
+\sum_{a=4}^7 V_K(r_{ij}) \lambda_i^a \lambda_j^a
+V_{\eta}(r_{ij}) \lambda_i^8 \lambda_j^8\right)
\vec\sigma_i\cdot\vec\sigma_j. \eqno(3.2)$$
The first term in (3.2) represents the pion-exchange interaction,
which acts only between
light quarks. The second term represents the kaon
exchange interaction,
which takes place in u-s and d-s pair states. The $\eta$-
exchange, which is represented by the third term, is allowed
in all quark pair states. In the $SU(3)_F$ symmetric limit
the constituent quark masses would be equal ($m_u = m_d = m_s$),
the pseudoscalar octet would be degenerate and the meson-constituent
quark coupling constant
would be flavor independent. In this limit the
form of the pseudoscalar exchange interaction reduces to (2.1),
which does not break the $SU(3)_F$ invariance of the baryon
spectrum. Beyond this limit the pion, kaon and $\eta$
exchange interactions will differ ($V_\pi \not= V_K \not= V_\eta$)
because of the difference between the strange and u, d quark
constituent masses ($m_{u,d} \not= m_s$), and because of the
mass splitting within the pseudoscalar octet
($\mu_\pi \not= \mu_K \not= \mu_\eta$) (and possibly also because
of flavor dependence in the meson-quark coupling constant).
 The source of both the $SU(3)_F$ symmetry
breaking constituent quark mass differences and the $SU(3)_F$ symmetry
breaking mass splitting of the pseudoscalar octet is
the explicit chiral symmetry breaking in QCD.\\

{\bf 4. The Structure of the Baryon Spectrum}

\vspace{0.5 cm}

 The  two-quark matrix elements of the
interaction (2.1) are:

$$<[f_{ij}]_F\times [f_{ij}]_S : [f_{ij}]_{FS}~| -V(r_{ij})
\vec \lambda^F_i \cdot \vec \lambda_j^F
\vec \sigma_i \cdot \vec \sigma_j ~|~[f_{ij}]_F
\times [f_{ij}]_S : [f_{ij}]_{FS}>$$
$$=\left\{\begin{array}{rr} -{4\over 3}V(r_{ij})& [2]_F,[2]_S:[2]_{FS} \\
-8V(r_{ij}) & [11]_F,[11]_S:[2]_{FS} \\
4V(r_{ij}) & [2]_F,[11]_S:[11]_{FS}\\ {8\over
3}V(r_{ij}) & [11]_F,[2]_S:[11]_{FS}\end{array}\right..\eqno(4.1)$$

\noindent
{}From these the following important properties may be inferred:

(i) At short range where $V(r_{ij})$ is positive the chiral
interaction (2.1) is attractive in the symmetrical FS pairs and
repulsive in the antisymmetrical ones. At large distances the potential
function $V(r_{ij})$ becomes negative and the situation is
reversed.

(ii) At short range  among the $FS$-symmetrical pairs
the flavor antisymmetrical pairs experience
a much larger attractive interaction than the flavor-symmetrical
ones and among the FS-antisymmetrical pairs
the strength of the repulsion in flavor-antisymmetrical
pairs is considerably weaker than in symmetrical ones.

\medskip
Given these properties we conclude that
with the given flavor symmetry the more symmetrical $FS$
Young pattern for a baryon - the more attractive contribution at short
range comes from the interaction (2.1). With two identical
flavor-spin Young patterns $[f]_{FS}$ the attractive contribution
at short range is larger in the case with the more antisymmetrical
flavor Young pattern $[f]_F$.

\medskip
Thus the $[3]_{FS}$ state in the $N(1440)$, $\Delta(1600)$
and
$\Sigma(1660)$ positive parity resonances from the $N=2$ band feels a
much stronger
attractive interaction than the mixed symmetry state $[21]_{FS}$ in the
$N(1535)$,
$\Delta(1700)$
and $\sum(1750)$ resonances ($N=1$ shell). Consequently the masses of the
positive parity states $N(1440)$, $\Delta(1600)$  and
$\Sigma(1660)$ are shifted
down relative to the other ones, which explains the reversal of
the otherwise expected "normal ordering".
The situation is different in the case of the $\Lambda(1405)$ and
$\Lambda(1600)$, as the flavor state of the $\Lambda(1405)$ is
totally antisymmetric. Because of this the
$\Lambda(1405)$ gains an
attractive energy, which is
comparable to that of the $\Lambda(1600)$, and thus the ordering
suggested by the confining oscillator interaction is maintained.

\medskip
If the confining interaction in each quark pair
is taken to have the harmonic oscillator form,
the exact eigenvalues and eigenstates to
the coinfining 3q Hamiltonian
 are

$$E=(N+3)\hbar\omega+3V_0,\eqno(4.2)$$
$$\Psi=|N(\lambda\mu)L[f]_X[f]_{FS}[f]_F[f]_S>,\eqno(4.3)$$

\noindent
where $N$ is the number of quanta in the state, the Elliott symbol
$(\lambda \mu)$ characterizes the $SU(3)$ harmonic oscillator symmetry,
and $L$ is the orbital momentum. The spatial ($X$), flavor-spin ($FS$),
flavor ($F$), and spin ($S$) permutational symmetries are indicated
by corresponding Young patterns (diagrams) $[f]$. All these functions are well
known and can be found e.g. in \cite{GLKU}.
Note that the totally antisymmetric color state
$[111]_C$, which is common to all the states, has been
suppressed in (4.3). By the Pauli principle $[f]_X=[f]_{FS}$.

The full Hamiltonian is the sum of the confining Hamiltonian
and the chiral field interaction (2.1).
When the
boson exchange interaction (2.1) is treated in first order perturbation
theory the mass of the baryon states takes the form

$$M=M_0+N\hbar\omega+ \delta M_\chi, \eqno(4.4)$$

\noindent
where the chiral interaction contribution is
$ \delta M_\chi = <\Psi|H_\chi|\Psi>,$
and
$M_0 = \sum_{i=1}^3 {m_i} + 3(V_0 + \hbar \omega).$
The chiral interaction contribution for each baryon
is a linear combination of the matrix elements of
the two-body potential $V(r_{12})$, defined as
$P_{nl}=<\varphi_{nlm}(\vec r_{12})
|V(r_{12})|\varphi_{nlm}(\vec r_{12})>.$\\

\noindent
{\small
\hspace{1.cm}{\bf Table 1.} The structure of the $\Lambda$-hyperon
states up to $N=2$, including
 predicted

\noindent
\hspace{1.cm} unobserved or nonconfirmed states indicated
by question marks. The predicted

\noindent
\hspace{1.cm}energies (in MeV)
are given in the brackets under the empirical values.}
{\footnotesize
\begin{center}
\begin{tabular}{|llll|} \hline
$N(\lambda\mu)L[f]_X[f]_{FS}[f]_F[f]_S$
& LS multiplet & average &$\delta M_\chi$\\
&&energy&\\ \hline
$0(00)0[3]_X[3]_{FS}[21]_F[21]_S$ & ${1\over 2}^+, \Lambda$ &
1115&$-14 P_{00}$\\
&&&\\
$1(10)1[21]_X[21]_{FS}[111]_F[21]_S$ & ${1\over 2}^-, \Lambda(1405);
{3\over 2}^-,\Lambda(1520)$ &
1462&$-12 P_{00}+4P_{11}$\\
&&(1512)&\\
$2(20)0[3]_X[3]_{FS}[21]_F[21]_S$ & ${1\over 2}^+, \Lambda(1600)$ &
1600&$-7 P_{00}-7P_{20}$\\
&&(1616)&\\
$1(10)1[21]_X[21]_{FS}[21]_F[21]_S$ & ${1\over 2}^-, \Lambda(1670);
{3\over 2}^-, \Lambda(1690)$ &
1680&$-7 P_{00}+5 P_{11}$\\
&&(1703)&\\
$1(10)1[21]_X[21]_{FS}[21]_F[3]_S$ & ${1\over 2}^-, \Lambda(1800);
{3\over 2}^-,\Lambda(?);$ &
1815&$-2 P_{00}+4P_{11}$\\
&${5\over 2}^-,\Lambda(1830)$&(1805)&\\

&&&\\
$2(20)0[21]_X[21]_{FS}[111]_F[21]_S$ & ${1\over 2}^+, \Lambda(1810)
$&1810&$-6P_{00}-6P_{20}+4P_{11}$\\
&&(1829)&\\
$2(20)2[3]_X[3]_{FS}[21]_F[21]_S$ & ${3\over 2}^+, \Lambda(1890);
{5\over 2}^+,\Lambda(1820)$ &
1855&$-7 P_{00}-7P_{22}$\\
&&(1878)&\\
$2(20)0[21]_X[21]_{FS}[21]_F[21]_S$&${1\over 2}^+,\Lambda(?)$&
?&$-{7\over 2}P_{00}-{7\over 2}P_{20}+5P_{11}$\\
&&(1954)&\\
$2(20)0[21]_X[21]_{FS}[21]_F[3]_S$ & ${3\over 2}^+, \Lambda(?)$&
?&$-P_{00}-P_{20}+4P_{11}$\\
&&(1989)&\\
$2(20)2[21]_X[21]_{FS}[21]_F[3]_S$ & ${1\over 2}^+, \Lambda(?);
{3\over 2}^+,\Lambda (?);$&2020?&
$-P_{00}-P_{22}+4P_{11}$\\
&${5\over 2}^+\Lambda(?);{7\over 2}^+,\Lambda(2020?)$&(2026)&\\
&&&\\
$2(20)2[21]_X[21]_{FS}[111]_F[21]_S$ & ${3\over 2}^+, \Lambda(?);
{5\over 2}^+,\Lambda(?)$&
?&$-6P_{00}-6P_{22}+4P_{11}$\\
&&(2053)&\\
$2(20)2[21]_X[21]_{FS}[21]_F[21]_S$ & ${3\over 2}^+,\Lambda(?);
{5\over 2}^+,\Lambda(2110)$ &2110?
&$-{7\over 2}P_{00}-{7\over 2}P_{22}+5P_{11}$\\
&&(2085)&\\ \hline
\end{tabular}
\end{center}
}

\noindent
Here $\varphi_{nlm}(\vec r_{12})$
represents the oscillator wavefunction
with n excited quanta.
 As we shall only consider
the baryon states in the $N\le 2$ bands we shall only need
the 4 radial matrix elements $P_{00},P_{11},P_{20}$ and
$P_{22}$ for the numerical construction of the spectrum.
In this approximate $SU(3)_F$-invariant  version of the chiral boson
exchange interaction the $\Lambda-N$ and the $\Xi - \Sigma$
mass differences
would solely be ascribed the  mass difference
between the s and u,d quarks since all these baryons have identical
orbital structure and permutational symmetries
and the states in the $\Lambda$-spectrum would be degenerate
with the corresponding states in the $\Sigma$-spectrum which have
equal symmetries.

\medskip

The  oscillator parameter $\hbar\omega$ and the 4 integrals
 are extracted from
the mass differences between the nucleon and the $\Delta(1232)$,
the $\Delta(1600)$ and
the $N(1440)$, as well as the splittings between the nucleon
and the average mass of the
two pairs of states $N(1535)-N(1520)$ and $N(1720)-N(1680)$.
This procedure yields the parameter values
$\hbar\omega$=157.4 MeV,
$P_{00}$=29.3 MeV, $P_{11}$=45.2 MeV, $P_{20}$=2.7 MeV and
$P_{22}$=--34.7 MeV. Given these values all other excitation energies
(i.e. differences between the masses of given resonances and
the corresponding ground states)
of the nucleon, $\Delta$- and $\Lambda$-hyperon spectra are
predicted to within
 $\sim$ 15\% of the empirical values
where known, and well within the uncertainty limits
of those values.
Note that these matrix elements provide a quantitatively satisfactory
description of the $\Lambda$-spectrum (see Table 1)
even though they are extracted from the $N-\Delta$ spectrum.

\medskip
The relative magnitudes and signs
of the numerical parameter values can be readily understood. If
the potential function $V(\vec r)$ is assumed to have the
form of a Yukawa function with a smeared $\delta$-function
term that is positive  at short range $r\le 0.6-0.7$  fm,
as suggested by the pion size $\sqrt{<r_\pi^2>}=0.66$ fm,
one expects $P_{20}$
to be considerably smaller than $P_{00}$ and $P_{11}$,
as the radial wavefunction for the excited S-state has a node,
and as it extends further into region of where the potential
is negative.
The negative value for $P_{22}$ is also natural as the
corresponding wavefunction is suppressed at short range
and extends well beyond the expected
0 in the potential function.
The relatively small value of the oscillator parameter (157.4 MeV)
leads to the empirical value 0.86 fm for the
nucleon radius $\sqrt{<r^2>}=\sqrt{\hbar/m\omega}$
if the light quark constituent mass is taken to be 330-340 MeV,
as suggested by the magnetic moments of
the nucleon.\\

{\bf 5. The  ${\bf SU(3)_F}$ Breaking Chiral Boson
Interaction}

\vspace{0.5cm}

The model described above has relied on an interaction potential function
$V(r)$ in (2.1) that is flavor independent. A refined version takes
into account the explicit flavor dependence of the potential function
in (3.2) ($V_\pi \not= V_K \not= V_\eta$). In the following I
shall show how this explicit flavor dependence provides us with an
explanation of the mass spliting between the $\Lambda$ and the $\Sigma$ which
have the same quark content and the same $FS$, $F$ and $S$
symmetries, i.e. they are degenerate within the $SU(3)_F$ version
(2.1) of the chiral boson exchange interaction.

\medskip
Beyond the $SU(3)_F$ limit the ground state baryons will be determined
by the $\pi$-exchange radial integral $P_{00}^\pi$, the $K$-exchange one,
$P_{00}^K$, and by the $\eta$-exchange integrals,
$P_{00}^{uu} = P_{00}^{ud} = P_{00}^{dd}$, $ P_{00}^{us}$ and $ P_{00}^{ss}$,
where the superscripts indicate quark pairs to which the $\eta$-exchange
applies. As indicated by the Yukawa
interaction (3.1) these matrix elements should be inversely
proportional to the product of the quark masses of the pair state. Thus
$P_{nl}^{us}={m_u\over m_s}P_{nl}^{uu},\quad P_{nl}^{ss}=({m_u\over
m_s})^2P_{nl}^{uu}.$
We also assume that $P_{00}^{us}\simeq P_{00}^{K}$, which is
suggested by the fact that the quark masses are equal in the states,
in which these interactions act, and by the near equality of the kaon
and $\eta$ masses, $\mu_\eta \simeq \mu_K$. Thus we have only two
independent radial integrals.

\medskip
 To determine
the integrals $P_{00}^\pi$, $P_{00}^K$ and the quark mass difference $\Delta_q
=m_s - m_u$ we consider the $\Sigma(1385)-\Sigma$,
$N-\Delta$ and $\Lambda - N$ splittings:

$$m_{\Sigma (1385)} - m_\Sigma = 4P_{00}^{us} + 6P_{00}^K, \eqno(5.1)$$
$$m_\Delta - m_N = 12P_{00}^\pi - 2P_{00}^{uu}, \eqno(5.2)$$
$$m_\Lambda - m_N = 6P_{00}^\pi - 6P_{00}^K + \Delta_q, \eqno(5.3)$$

\noindent
which imply
$P_{00}^{K}$ = 19.6 MeV,
$\Delta_q=121$ MeV if the conventional value 340 MeV is given to
$m_u$, $P_{00}^{\pi} =
28.9$ MeV and the quark mass ratio
$m_s/m_u=1.36$. These matrix element values   lead to the
values 65 MeV and 139 MeV for the $\Sigma-\Lambda$ and the $\Xi
-\Sigma$ mass differences

$$m_\Sigma - m_\Lambda = 8P_{00}^\pi-4P_{00}^K - \frac {4}{3}P_{00}^{uu}
-\frac {8}{3}P_{00}^{us}, \eqno(5.4)$$
$$m_\Xi - m_\Sigma = P_{00}^\pi + \frac {1}{3}P_{00}^{uu}
-\frac {4}{3}P_{00}^{ss} + \Delta_q. \eqno(5.5)$$

\noindent
  in good agreement
with the empirical values 77 MeV and 125 MeV respectively.\\

\vspace{0cm}

{\bf 6. Exchange Current Corrections to the Magnetic Moments}
\vspace{0.5cm}

A flavor dependent interaction of the form (2.1) will imply the
presence of an irreducible two-body exchange current
operator, as seen e.g. directly from the continuity equation,
by which the commutator of the interaction and the single
particle charge operator equals the divergence of the exchange
current density \cite{RISK}. Because this commutator vanishes
with interparticle separation  this exchange current
is however a priori expected to be of less importance for baryons,
than for nuclei, in which the longer range of the wave functions
can lead to large matrix elements of the pion exchange current
operator.
This is one contributing reason for why the naive constituent quark
model provides such a successful description of the magnetic moments.

\medskip
The general form of the octet vector exchange current operator
that is associated
with the complete octet mediated interaction (3.2) will have the form
\cite{GLO2}

$$\vec \mu^{ex}=\mu_N\{\tilde{V}_\pi(r_{ij})(\lambda_i^1\lambda_j^2-\lambda_i^2
\lambda_j^1)
+\tilde{V}_K(r_{ij})(\lambda_i^4\lambda_j^5-\lambda_i^5
\lambda_j^4)\}(\vec\sigma_i \times \vec \sigma_j).\eqno(6.1)$$

We find \cite{GLO2} that the meson exchange current contributions
systematically
improve predictions of the naive constituent quark model (i.e. with
one-body quark currents only) for all known magnetic moments.
However these contributions are not
large and do not exceed 10\% in agreement with the expectation above.\\

{\bf 7. Resolution of the ${\bf N^* \rightarrow N\eta}$ puzzle}

\vspace {0.5cm}

I will suggest here a simple explanation for a strong selectivity
of the $N\eta$ ($\Lambda \eta$ and $\Sigma \eta$) decay branching
ratios of the $N^*$ ($\Lambda^*$ and $\Sigma^*$) baryons and at the
same time an absence of such a selectivity for the $N\pi$ ($NK$)
decays within the chiral quark model outlined above \cite{GLO3}.

\medskip
It is well seen from the matrix elements (4.1) that at short range there is
 very strong attraction between quarks  in the $S_{ij}=T_{ij}=0$
and  strong repulsion in the $S_{ij}=1,T_{ij}=0$ and $S_{ij}=0,T_{ij}=1$
channels. The chiral field interaction is attractive, but rather weak,
in the $S_{ij}=T_{ij}=1$ pair state. Thus those baryons which contain
the $S_{ij}=T_{ij}=0$ quark pair state will be strongly clusterized into
quark-diquark configuration with  $S_{12}=T_{12}=0$ diquark quantum numbers.
These will be all baryons with $[21]_F [21]_S$ symmetries of  zero
order wave function and with the smallest possible orbital momentum
$L$ =0 or 1 ($N$, $N(1440)$, $N(1535)$ and $N(1710)$). The baryons with
the  $[21]_F [3]_S$ symmetries (N(1650),...) have the quark pair components
$S_{ij}=T_{ij}=1$ and $S_{ij}=1,T_{ij}=0$. In the former there is rather
soft attraction and in the latter there is much larger repulsive
interaction at short range. This will again lead to a clusterization
into quark-diquark configuration but with  $S_{ij}=T_{ij}=1$ quantum
numbers for diquark. It is a clusterization into quark-diquark configurations
with different quantum numbers of diquarks in different baryons which
provides the sought explanation of the $\eta$ decay puzzle.

\medskip
 Consider first the case of the $N(1650)$.
As in $\eta$ decay the isospin of the involved quark
 is unchanged it cannot proceed through the
$S_{12}=1,T_{12}=1\rightarrow S_{12}=0,T_{12}=0$
transition that would connect the diquark states in the
$N(1650)$ and the nucleon.
The corresponding pion decay, in which isospin flip
is possible, can on the other hand connect these pair
states.
The reason for the large $\eta$ decay branch of the $N(1535)$ in contrast
is that its  wavefunction  has diquark with the
same spin-isospin structure as the nucleon.

\medskip
This argument generalizes to the predictions that:
(i) all baryon resonances above the corresponding $\eta$ decay threshold with
$[21]_{FS}[21]_F[21]_S$ symmetry
zero order wavefunctions and smallest possible
orbital and total angular momentum
($\frac {1}{2}^-, N(1535);$
$\frac {1}{2}^+, N(1710);$
$\frac {1}{2}^-, \Lambda(1670);$ $ \frac {1}{2}^-,
\Sigma(1750)$)
should have large $\eta$-decay branching
ratios whereas
(ii) the baryon resonances that have $[21]_{FS}[21]_F[3]_S$ symmetry
zero
order wave functions should have strongly suppressed
$\eta$-decay branching ratios.
This prediction is in excellent agreement with the
corresponding empirical branching ratios, all of which are
large for the baryons in the list (i) and vanish for all other baryons.

\medskip
\medskip
\medskip
Be it as it may, the present organization of fine
structure of the baryon
spectrum based on the quark-quark interaction that is
mediated by the octet of pseudoscalar mesons, which
represent the Goldstone bosons associated with the
hidden mode of the approximate chiral symmetry of QCD
is both simple and phenomenologically successful.
The predicted energies of the states in
the nucleon and strange hyperon spectra agree with the
empirical values, where known, to within a few percent.
 This interaction between
light and strange quarks inside the charm and bottom hyperons
with one heavy quark is also of crucial importance for those baryons
\cite{GLO4}.

\medskip
The very satisfactory predictions obtained here for the baryon
spectrum reconcile the quark model for baryons with the phenomenologically
successful meson exchange description of the nucleon-nucleon
interaction.

\medskip
I should also mention  that the coupling between the constituent
quarks and the pseudoscalar mesons governs the structure and content
of the quark sea. It has recently been shown to resolve well
known problems that are associated with the strangeness content
of the nucleon and the nucleon spin structure as measured in
deep inelastic lepton-proton scattering \cite{QUIGG,CHENG}.

{\small

\
}

\end{document}